\documentclass[12pt, nonatbib]{elsarticle}

\usepackage{amssymb}
\usepackage{amsmath}
\usepackage{caption}
\usepackage[hidelinks]{hyperref}
\usepackage{subcaption}
\usepackage{soul}
\usepackage{color}
\usepackage{setspace}
\renewcommand\hl[1]{#1}
\makeatletter 
\let\c@author\relax
\makeatother
\usepackage[backend=biber,style=chem-angew,hyperref=true, articletitle=true, doi=true]{biblatex}
\journal{Journal of Magnetic Resonance}

\begin{document}

\begin{frontmatter}

\title{Generalized Cross-Validation as a Method of Hyperparameter Search for MTGV Regularization}

\author[1,2,3]{Julian B. B. Beckmann \corref{cor1}}
\ead{jbbeckmann@mgh.harvard.edu}
\author[1]{Amy Sparks}
\author[1]{Jordan A. Ward-Williams}
\author[1]{Mick D. Mantle}
\author[1]{Andrew J. Sederman}
\author[1]{Lynn F. Gladden}

\affiliation[1]{organization={University of Cambridge, Department of Chemical Engineering and Biotechnology},
            addressline={Philippa Fawcett Drive}, 
            city={Cambridge},
            postcode={CB3 0AS}, 
            country={United Kingdom}}
\affiliation[2]{organization={Harvard Medical School},
            addressline={25 Shattuck Street}, 
            city={Boston},
            postcode={MA 02115}, 
            country={United States}}
\affiliation[3]{organization={Martinos Centre for Biomedical Imaging, Massachusetts General Hospital},
            addressline={149 13th Street}, 
            city={Charlestown},
            postcode={MA 02129}, 
            country={United States}}
            
\cortext[cor1]{Corresponding author}  

\begin{abstract}
The concept of generalized cross-validation (GCV) is applied to modified total generalized variation (MTGV) regularization. Current implementations of the MTGV regularization rely on manual (or semi-manual) hyperparameter optimization, which is both time-consuming and subject to bias. The combination of MTGV-regularization and GCV allows for a straightforward hyperparameter search during regularization. This significantly increases the efficiency of the MTGV-method, because it limits the number of hyperparameters, which have to be tested and, improves the practicality of MTGV regularization as a standard technique for inversion of NMR signals. The combined method is applied to simulated and experimental NMR data and  the resulting reconstructed distributions are presented. It is shown that for all data sets studied the proposed combination of MTGV and GCV minimizes the GCV score allowing an optimal hyperparameter choice.   
\end{abstract}


\end{frontmatter}

\section{Introduction}
\label{sec:Int}

The ability to measure spin lattice relaxation~$T_1$, spin spin relaxation~$T_2$ and diffusion~$D$ makes NMR \hl{experiments an} indispensable tool across a wide-ranging area of disciplines. For instance, in the field of medical imaging $T_1$, $T_2$ or $D$ are used to generate contrast between different types of tissue \supercite{Banerjee2014MultiparametricDisease, Kellman2014T1-mappingPrecision, Wood2014UseOverload, Wood2005MRIPatients, Galea2013LiverImaging, Partridge2013DiffusionApplications}. But also in the chemical industry, estimation of the latter physical parameters is used to study surface interactions or exchange in a multitude of different systems such as oil bearing rocks or process chemicals imbibed in catalytic pellets.\supercite{DAgostino2014InterpretationMaterials, Robinson2018DirectMaterial, DAgostino2016SolventMeasurements, Weber2010SurfaceCatalysts, Gladden2006MagneticProcesses, Kort2018AcquisitionMRI, Hertel2019FastCPMG} Despite the richness of information, which can be inferred already from a one-dimensional $T_1$-, $T_2$- or $D$-experiment, the physical interpretation of those measurements can be ambiguous due to the complexity of the acquired signal especially in multi-component systems. One solution to tackle this issue is to correlate $T_1$-, $T_2$- or $D$-measurements with each other. In a NMR experiment, this can be easily done in a \hl{consecutive} manner. For instance, a correlation between spin lattice and spin spin relaxation can be simply achieved by adding a one-shot CPMG acquire to a saturation or inversion recovery sequence.\supercite{Chandrasekera2008RapidCorrelations, Fleury2016CharacterizationMaps} In this example, the saturation or inversion recovery sequence encodes the $T_1$-dimension, whereas the CPMG acquire is responsible for the $T_2$-encoding, which finally provides the experimentalist with the data which after inversion yields a $T_1$-$T_2$-correlation map. This procedure can be used for any possible two-dimensional combination of $T_1$-, $T_2$- or $D$-encodings, but can also be expanded to three or even more dimensions. This leads to a plethora of possible correlation experiments and consequently, a wide range of applications across different disciplines is found. For example, $T_1$-$T_2$-experiments\supercite{Chandrasekera2008RapidCorrelations, Fleury2016CharacterizationMaps} are used to study surface interactions at liquid-solid interfaces, while $D$-$D$-\supercite{Neudert2011DiffusionFields, Callaghan2003UseMedia, Callaghan2004Diffusion-diffusionDynamics, Mankinen2020UltrafastResonance} and $T_2$-$T_2$-experiments\supercite{Song2016TheFeatures, Sun2011MeasurementSpectroscopy} are employed in probing exchange between different chemical or physical environments. Although, the experimental implementation of those experiments is straightforward, there is a major theoretical challenge to overcome, while processing the experimental data. In more detail, the signal acquired in the NMR experiment is considered to be a Laplace transform of the sought-after distribution.\supercite{PaulCallaghan2011TranslationalResonance} Due to the ill-conditioned nature of the Laplace transform no well-defined analytical inverse transform exists \hl{unlike} the case for the Fourier transform, which means that so-called inversion methods have to be employed to obtain the distribution from the NMR signal.\supercite{Mitchell2012NumericalDimensions} This problem has drawn a lot of attention across a multitude of different disciplines ranging from statistics\supercite{Hastie2009TheLearning, Gareth2021AnLearning} up to image processing.\supercite{Wen2018UsingProblems} Currently, the most prominent approach to tackle this issue is the application of Tikhonov regularization.\supercite{Venkataramanan2002SolvingDimensions, Song2002T1T2Inversion, Huber2017AnalysisApproximations, Medellin2016MultidimensionalInversion, Su2016AnL-curve, Guo2019NuclearFunctions} The main benefit of Tikhonov regularization lies in its stability regarding noise but at the same time, it suffers from the condition that the sought-after distribution has to be smooth.\supercite{Tikhonov1978SolutionsProblems., Fuhry2012AMethod, Golub1999TikhonovSquares} Therefore, Tikhonov regularization can only differentiate between features which differ at least by a factor of three in relaxation time or diffusion coefficient and furthermore, the experimentalist has to know a priori that the investigated distribution is expected to be smooth.\supercite{PaulCallaghan2011TranslationalResonance, Mitchell2012NumericalDimensions} Both issues were already addressed in earlier work. For instance, Reci \textit{et al.} showed that $L_1$-regularization can be successfully used to reconstruct sparse distributions, but at the same time, $L_1$-regularization suffers from poor performance with smooth distributions.\supercite{Reci2017RetainingExperiments.} In contrast, a compromise between sparsity and smoothness could be achieved with modified total generalized variation (MTGV), which has the further advantage that no prior knowledge regarding the sparsity or smoothness of the distribution is necessary.\supercite{Reci2017RetainingExperiments.}  Despite the promising results achieved using the MTGV method a challenge remains regarding the estimation of suitable hyperparameters. Specifically, the cost function of MTGV regularization employs two penalty terms. The first one ensures sufficient sparsity, whereas the second term enforces smooth features. The trade-off between fidelity, sparsity and smoothness is controlled by the choice of the hyperparameters $\alpha$ and $\beta$.\supercite{Reci2017RetainingExperiments.} In its current form, the MTGV algorithm relies on the experimentalist to choose appropriate hyperparameters, which can be very time consuming and presumably allows  for bias and potentially incorrect results and thus, limits the practicality of the method for routine usage significantly. Therefore, this paper aims for the implementation of an automated hyperparameter search for MTGV regularization. For Tikhonov regularization, a considerable number of methods for automated hyperparameter searches are employed in the literature including different scoring metrics such as leave-one-out cross-validation, generalized cross-validation, the L-curve method and the Butler-Reeds-Dawson method.\supercite{Mitchell2012NumericalDimensions, Orr1996IntroductionNetworks, Butler1981EstimatingSmoothing} In \hl{practice}, generalized cross-validation (GCV) is used predominantly due to its benefits regarding robustness, ease of implementation and convergence speed.\supercite{Golub1979GeneralizedParameter, Bottegal2018TheFilter, Whatmough1994ApplyingRestoration} Those advantages are also the reason, that GCV was chosen as one the scoring metrics in this paper. However, the combination of MTGV regularization with GCV is not \hl{straightforward}, because no closed-form expression of the  GCV score can be inferred from the cost function of the MTGV minimisation problem.\supercite{Orr1996IntroductionNetworks, Wen2018UsingProblems} Recently, similar issues started to draw more attention in the mathematical community and Wen \textit{et al.} were able to prove that for the example of total variation regularization the cost function can be rearranged to a Tikhonov regularization problem for which a  GCV score is easily computable.\supercite{Wen2018UsingProblems} Hence, in this paper, it will be shown that this approach is also valid for MTGV regularization and in consequence, a method to calculate a  GCV score for the MTGV regularization problem will be derived. Finally, the  GCV score can be minimized and the best regularization parameters can be chosen automatically.

\section{Theoretical Background}
\label{sec:Theo}

In the field of NMR, the signal of a correlation or exchange experiment is given by the Laplace transform of some physical distribution. Consequently, in mathematical terms the NMR signal acquired is given as follows:\supercite{Mitchell2012NumericalDimensions, Reci2017ObtainingProblems, Reci2017RetainingExperiments., PaulCallaghan2011TranslationalResonance}
\begin{equation} \label{eq:Sig_Math}
    \begin{split}
        S(x_1, x_2,...) = & \int_{0}^{\infty} \int_{0}^{\infty} F(X_1, X_2,...) K_1(x_1, X_1) K_2(x_2, X_2)... \; dX_1 \, dX_2... \\
        & + E(x_1, x_2,...),
        \end{split}
\end{equation}
where $F$ is the distribution of some variables $X_{1,2,...}$, $K_{1,2,...}$ the kernel functions, $E$ the inherent noise in the signal and $x_{1,2,...}$ are the variables used to encode for $X_{1,2,...}$ in the NMR pulse sequence. Common examples for $X_{1,2,...}$ are the diffusion coefficient $D$ and relaxation time constants $T_{1,2}$, which means that $x_{1,2,...}$ usually coincides with a magnetic field gradient $\mathbf{g}$ or a time delay $\tau$.\supercite{Mitchell2012NumericalDimensions, PaulCallaghan2011TranslationalResonance} The kernel functions $K_{1,2,...}$ are given in the most cases as some sort of exponential decay, where the exact form depends on the pulse sequence used for encoding.\supercite{Mitchell2012NumericalDimensions, PaulCallaghan2011TranslationalResonance} Equation \ref{eq:Sig_Math} can be re-written into a discrete form, which reads as follows:
\begin{equation} \label{eq:SigDis}
    \underline{\mathbf{S}} = \underline{\mathbf{K}}\,\underline{\mathbf{F}} + \underline{\mathbf{E}},
\end{equation}
where $\underline{\mathbf{K}}$ is the kernel matrix, $\underline{\mathbf{S}}$ is the discretized vector of $S$ and analogously, the same holds true for $\underline{\mathbf{F}}$ and $\underline{\mathbf{E}}$. Hence, the inversion problem which has to be solved is to estimate the distribution~$\underline{\mathbf{F}}$, while the kernel~$\underline{\mathbf{K}}$ and the NMR signal~$\underline{\mathbf{S}}$ are known. In the case of the MTGV regularization this leads to the following minimisation problem:\supercite{Reci2017RetainingExperiments.}
\begin{equation} \label{eq:MTGV}
    \underline{\mathbf{F}} = \arg \; \min {}_{\underline{\mathbf{F}} \, \geq \, 0, \underline{\mathbf{W}}} \left(\frac{\alpha}{2} ||\underline{\mathbf{K}} \, \underline{\mathbf{F}} - \underline{\mathbf{S}}||_2^2 + ||\underline{\mathbf{F}} - \underline{\mathbf{W}}||_1 + \beta ||\underline{\mathbf{D}}_{\mathbf{2}} \, \underline{\mathbf{W}}||_1^* \right),
\end{equation}
where $\underline{\mathbf{W}}$ is an auxiliary vector, $\underline{\mathbf{D}}_{\mathbf{2}}$ is the matrix which performs a second derivative of the vector it is applied on. The norm $||...||_p$ refers to the $\mathrm{L}_p$-norm, which is given as follows:\supercite{Reci2017RetainingExperiments.}
\begin{equation} \label{eq:Norm}
    \mathrm{L}_p \left(\underline{\mathbf{A}} \right) = ||\underline{\mathbf{A}}||_p = \left(\sum_{i = 1}^n |a_i|^p \right)^{\frac{1}{p}},
\end{equation} 
where $\underline{\mathbf{A}}$ refers to any real and finite vector with $a_1$ to $a_n$ entries and p is a real number. The definition of $||...||_1^*$ differs slightly, because the second derivative of a matrix can be calculated in several directions. The exact definitions of $\underline{\mathbf{D}}_{\mathbf{2}}$ and $||...||_1^*$ can be found in the original work of Reci and co-workers.\supercite{Reci2017RetainingExperiments.} In reference to equation \ref{eq:MTGV}, it can be seen that the \hl{MTGV} cost function consist of two penalty terms. The first one ensures sufficient sparsity, whereas the second enforces smooth features and the balance between both is determined via the hyperparameter $\beta$. Thus, MTGV can offer a compromise between sparsity and smoothness, which cannot be achieved with Tikhonov- or $\mathrm{L_1}$-regularization on their own.\supercite{Reci2017RetainingExperiments.} Additionally, the trade-off between fidelity and regularization can be controlled via the hyperparameter~$\alpha$. Hence, the selection of both hyperparameters  strongly influences the result of the minimisation problem, which is given by equation~\ref{eq:MTGV} and subsequently, the distributions which are finally reconstructed. Thus, the choice of hyperparameters is crucial for obtaining distributions, which actually represent physical reality. 
\section{Proposed Method}
\label{sec:Meth}

From Reci \textit{et al.}, it is known, that the minimization problem in equation~\ref{eq:MTGV} can be reformulated to a minimax problem, which reads as follows:\supercite{Reci2017RetainingExperiments., Rockafellar1970ConvexAnalysis, Chambolle2011AImaging}
\begin{equation} \label{eq:MiniMax}
\begin{split}
    \left(\underline{\mathbf{F}},\underline{\mathbf{W}},\underline{\mathbf{Y}}_{\mathbf{1}},\underline{\mathbf{Y}}_{\mathbf{2}} \right) = & \arg \; \min {}_{\underline{\mathbf{F}} \, \geq \, 0, \underline{\mathbf{W}}} \; \max {}_{\underline{\mathbf{Y}}_{\mathbf{1}},\underline{\mathbf{Y}}_{\mathbf{2}}} \left(\frac{\alpha}{2} ||\underline{\mathbf{K}} \, \underline{\mathbf{F}} - \underline{\mathbf{S}}||_2^2 \right. \\ & + \underline{\mathbf{Y}}_{\mathbf{1}}^{\mathbf{T}} \left(\underline{\mathbf{F}} - \underline{\mathbf{W}}  \right) + \underline{\mathbf{Y}}_{\mathbf{2}}^{\mathbf{T}} \, \underline{\mathbf{D}}_{\mathbf{2}} \, \underline{\mathbf{W}} \\ & \left. - h\left(\underline{\mathbf{Y}}_{\mathbf{1}} \right) - h\left(\underline{\mathbf{Y}}_{\mathbf{2}}/\beta \right) \vphantom{\frac{\alpha}{2}} \right),
\end{split}
\end{equation}
where $\underline{\mathbf{Y}}_{\mathbf{1,2}}$ are auxiliary vectors and $h$ is the indicator function defined as:
\begin{equation} \label{eq:Indic}
    h\left(\underline{\mathbf{Y}}_{\mathbf{1}} \right) = 
    \begin{cases}
        0, & ||\underline{\mathbf{Y}}_{\mathbf{1}}||_{\infty} \leq 1 \\
        \infty, & ||\underline{\mathbf{Y}}_{\mathbf{1}}||_{\infty} > 1
    \end{cases} ,
\end{equation}
where $||...||_{\infty}$ relates to $\lim_{\,p \, \to \, \infty} \, \mathrm{L}_p$. Equation \ref{eq:MiniMax} describes a primal-dual problem, which was tackled in Reci's work by the primal-dual hybrid gradient method. The full iteration scheme of the MTGV algorihtm can be found in the original MTGV paper of Reci \textit{et al.},\supercite{Reci2017RetainingExperiments.} but at this point only the update formula of the distribution $\underline{\mathbf{F}}$ is needed, which is given by the following expression:\supercite{Reci2017RetainingExperiments., Chambolle2011AImaging}
\begin{equation} \label{eq:Fit}
    \underline{\mathbf{F}}^{\mathbf{k}+ 1} = \left(\underline{\mathbf{I}} + \tau \alpha \, \underline{\mathbf{K}}^{\mathbf{T}} \, \underline{\mathbf{K}} \right)^{-1} \left(\underline{\mathbf{F}}^{\mathbf{k}} - \tau \, \underline{\mathbf{Y}}_{\mathbf{1}}^{\mathbf{k} + 1} + \tau \alpha \, \underline{\mathbf{K}}^{\mathbf{T}} \, \underline{\mathbf{S}} \right),
\end{equation}
where $\mathbf{k}$ is an integer, which counts the number of iterations, $\underline{\mathbf{I}}$ is the identity matrix and $\tau$ refers to a constant controlling the convergence of the algorithm. Equation \ref{eq:Fit} can be rearranged into: \supercite{Wen2018UsingProblems}
\begin{equation} \label{eq:FitTik}
    \underline{\mathbf{F}}^{\mathbf{k} + 1} = \underline{\mathbf{U}}^{\mathbf{k} + 1} + \underline{\mathbf{F}}^{\mathbf{k}} - \tau \, \underline{\mathbf{Y}}_{\mathbf{1}}^{\mathbf{k} + 1},
\end{equation}
with
\begin{equation} \label{eq:U}
\begin{split}
    \underline{\mathbf{U}}^{\mathbf{k} + 1} = & \arg \; \min {}_{\underline{\mathbf{U}} \, \geq \, \tau \, \underline{\mathbf{Y}}_{\mathbf{1}}^{\mathbf{k}} - \, \underline{\mathbf{F}}^{\mathbf{k}}} \left(\frac{1}{2 \tau \alpha} ||\underline{\mathbf{U}} ||_2^2 \right. \\ & \left. + \frac{1}{2} ||\underline{\mathbf{K}} \, \underline{\mathbf{U}} - \left(\underline{\mathbf{S}} - \underline{\mathbf{K}} \left(\underline{\mathbf{F}}^{\mathbf{k}} - \tau \, \underline{\mathbf{Y}}_{\mathbf{1}}^{\mathbf{k} + 1} \right) \right) ||_2^2 \vphantom{\frac{1}{2 \tau \alpha}} \right).
\end{split}    
\end{equation}
Equation \ref{eq:U} describes a Tikhonov-regularization problem, which can be solved in Matlab with standard optimisation functions such as fminunc and a GCV score for this can be easily calculated. The GCV equation is defined as follows:\supercite{Mitchell2012NumericalDimensions, Orr1996IntroductionNetworks, Wen2018UsingProblems} 
\begin{equation} \label{eq:GCV_mtgv}
\begin{split}
     & \mathrm{GCV} \left(\alpha \right) = \, \mathrm{Tr} \left(\underline{\mathbf{I}} \right) \\ & \times||\left(\underline{\mathbf{I}} - \underline{\mathbf{K}} \left(\underline{\mathbf{K}}^{\mathbf{T}} \, \underline{\mathbf{K}} + \frac{1}{\tau \alpha} \, \underline{\mathbf{I}} \right)^{-1} \underline{\mathbf{K}}^{\mathbf{T}} \right) \left(\underline{\mathbf{S}} - \underline{\mathbf{K}} \left(\underline{\mathbf{F}}^{\mathbf{k}} - \tau \, \underline{\mathbf{Y}}_{\mathbf{1}}^{\mathbf{k} + 1} \right) \right)||_2^2 \\ & \times \left(\mathrm{Tr} \left(\underline{\mathbf{I}} - \underline{\mathbf{K}} \left(\underline{\mathbf{K}}^{\mathbf{T}} \, \underline{\mathbf{K}} + \frac{1}{\tau \alpha} \, \underline{\mathbf{I}} \right)^{-1} \underline{\mathbf{K}}^{\mathbf{T}} \right) \right)^{-2},  
\end{split}
\end{equation}
where $\mathrm{Tr}$ refers to the trace of a matrix. Consequently, $\alpha$ should be chosen to minimize the latter equation. However, from equation~\ref{eq:GCV_mtgv} it is possible to derive an update formula for $\alpha$, which can be used in an iterative manner to estimate an optimal value for $\alpha$. The update formula is given by the following expression:\supercite{Orr1996IntroductionNetworks}
\begin{equation} \label{eq:GCVupdate_mtgv}
\begin{split}
    \alpha^{\mathrm{k} + 1} = & \left( \vphantom{\left(\underline{\mathbf{K}}^{\mathbf{T}} \, \underline{\mathbf{K}} + \frac{1}{\tau \alpha^{\mathrm{k}}} \, \underline{\mathbf{I}} \right)^{-1}} \tau \, \mathrm{GCV} \left(\alpha^{\mathrm{k}} \right) \; \mathrm{Tr} \left(\underline{\mathbf{I}} - \underline{\mathbf{K}} \left(\underline{\mathbf{K}}^{\mathbf{T}} \, \underline{\mathbf{K}} + \frac{1}{\tau \alpha^{(\mathrm{k})}} \, \underline{\mathbf{I}} \right)^{-1} \underline{\mathbf{K}}^{\mathbf{T}} \right) \right. \\ & \left. \times \, \mathrm{Tr} \left(\left(\underline{\mathbf{K}}^{\mathbf{T}} \, \underline{\mathbf{K}} + \frac{1}{\tau \alpha^{\mathrm{k}}} \, \underline{\mathbf{I}} \right)^{-1} - \frac{1}{\tau \alpha^{\mathrm{k}}} \left(\underline{\mathbf{K}}^{\mathbf{T}} \, \underline{\mathbf{K}} + \frac{1}{\tau \alpha^{\mathrm{k}}} \, \underline{\mathbf{I}} \right)^{-2} \right) \right)^{-1} \\ & \times \underline{\mathbf{U}}^{\mathbf{(k + 1),T}} \left(\underline{\mathbf{K}}^{\mathbf{T}} \, \underline{\mathbf{K}} + \frac{1}{\tau \alpha^{\mathrm{k}}} \, \underline{\mathbf{I}} \right)^{-1} \underline{\mathbf{U}}^{\mathbf{(k + 1)}}.
\end{split}
\end{equation}
After an initial $\alpha$ is chosen, equation~\ref{eq:GCV_mtgv} and~\ref{eq:GCVupdate_mtgv} can be used to calculate a GCV score and subsequently, to update $\alpha$ iteratively until a convergence criterion is met. This means that the primal dual algorithm has to be employed twice. Firstly, to estimate $\alpha$ and in a second instance, to reconstruct the distribution $\underline{\mathbf{F}}$ with the obtained $\alpha$. The latter procedure provides a rational for choosing $\alpha$, but to my knowledge, it is not possible to employ a similar GCV method for the selection of $\beta$. Conceptually the simplest way to select $\beta$ would be looping through a list of $\beta$'s and selecting the one, which minimizes the scoring-metric in use. However, with this method it is often necessary to explore a vast number of $\beta$'s rendering the overall inversion algorithm inefficient and it further imposes an additional bias due to the pre-selection of the $\beta$-list. Clearly, a more efficient way is to use another selection method, which chooses $\beta$ based on mathematical or statistical principles. Due to the non-availability of GCV or similar methods, the Butler-Reeds-Dawson (BRD) method was used to optimise $\beta$. In this case, the benefits of BRD are its ease of implementation, its computational efficiency and that no further time-consuming calculations of additional quantities are necessary. The BRD score is given through the following expression:\supercite{Mitchell2012NumericalDimensions, Butler1981EstimatingSmoothing}
\begin{equation} \label{eq:BRD}
    \mathrm{BRD} = \frac{||\underline{\mathbf{K}} \, \underline{\mathbf{F}} - \underline{\mathbf{S}}||_2}{\sigma},
\end{equation}
where $\sigma$ is the noise of the signal vector $\underline{\mathbf{S}}$. The update formula is defined as follows: \supercite{Mitchell2012NumericalDimensions, Butler1981EstimatingSmoothing}
\begin{equation} \label{eq:BRDupdate}
    \beta^{\mathrm{k + 1}} = \frac{\sqrt{\mathrm{Tr} \left(\underline{\mathbf{I}} \right)} \, \beta^{\mathrm{k}}}{||\underline{\mathbf{K}} \, \underline{\mathbf{F}} - \underline{\mathbf{S}}||_2}
\end{equation}
Hence, the final method employs the following iteration scheme. After initial values for $\alpha$ and $\beta$ are chosen, $\alpha$ is optimised as outlined in this section and subsequently, a distribution is reconstructed, but it should be remembered that at this stage, the initial $\beta$ is used for reconstruction. Consequently, in the next step $\beta$ is updated via the BRD method, which marks the end of one iteration cycle. The latter sequence is repeated until the BRD stopping criterion is fulfilled and the final distribution reconstructed.
\section{Simulations and Experiments}
\label{sec:ExSim}

This publication employs 2D-NMR signals derived from simulation as well as experiments. The benefit of simulated data is that the real distribution is known, which allows the real distribution to be compared with the reconstructed one. Furthermore, the signal-to-noise ratio can be freely adjusted in a simulation and consequently, the robustness of the developed method to noise can be tested. However, emulating experimental inaccuracies and signal imperfections are difficult to include in a simulation and therefore, their effect on the reconstruction can only be validated if experimental data is included. To simulate the NMR signal, equation \ref{eq:SigDis} is used, where $\underline{\mathbf{F}}$ is assumed to be a superposition of log-normal distributions with unconstrained mean and variance, whereas $\underline{\mathbf{E}}$ is chosen to achieve a certain signal-to-noise ratio. Overall, three type of experiments were simulated: $T_1\mbox{-}D$, $D\mbox{-}T_2$ and $T_1\mbox{-}T_2$. Each data set spans $32$ times $32$ data points and the signal-to-noise ratio was set to $1000$. The same type of experimental data was generated from standard $T_1\mbox{-}D$-, $D\mbox{-}T_2$- and $T_1\mbox{-}T_2$-experiments. The size of the experimental data sets varies between $16$ times $255$ data points for the $D\mbox{-}T_2$- or the $T_1\mbox{-}T_2$-experiments and $16$ times $16$ data points for the $T_1\mbox{-}D$-experiment. The signal-to-noise ratio varied depending on the type of experiment, but overall a minimum signal-to-noise ratio of $150$ was ensured. Prior to regularization, all data sets were truncated with the standard technique as described by Venkataramanan~\textit{et~al.}.\supercite{Venkataramanan2002SolvingDimensions, Song2002T1T2Inversion} In order to reduce, the time required to find an optimal $\alpha$, an initial $\alpha$ was chosen via the following equation:\supercite{Mitchell2012NumericalDimensions}
\begin{equation} \label{eq:AlphaInit}
    \alpha = \frac{\mathrm{Tr} \left(\underline{\mathbf{I}} \right)}{\mathrm{Tr} \left(\underline{\mathbf{K}}^{\mathbf{T}} \, \underline{\mathbf{K}} \right)}
\end{equation}
The rational behind this initialisation method, which is set to underestimate the optimal $\alpha$ but not drastically, is to reduce the number of inversions with very small values for $\alpha$ and consequently, to improve the hyperparameter search efficiency.\supercite{Mitchell2012NumericalDimensions} On the contrary, an initial $\beta$ was chosen empirically. Within the scope of this study, $10^{-10}$ as an initial $\beta$ provides a compromise between range and speed of the hyperparameter search. In more detail, starting at $10^{-10}$ resulted in fewer than twelve explored $\beta$-values for every data set tested in this paper. Further, including small values for $\beta$ in the hyperparameter search also ensures, that possible sparse solutions to the regularization problem are not missed. In terms of hyperparameter selection, the following procedures were employed. In the case of the $\alpha$-search, the $\alpha$ which fulfilled the convergence criterion was chosen, whereas for the choice of $\beta$ a more elaborated scheme based on the heel-criterion was used.\supercite{Venkataramanan2002SolvingDimensions, Song2002T1T2Inversion} This selects $\beta$ at the heel of the BRD-curve, which allows for multiple sets of hyperparameters, because the exact position of the heel is usually not clearly defined. Consequently, this method can incorporate different level of smoothness and sparsity depending on the requirements for the reconstructed distribution.
\section{Results and Discussion} \label{sec:Res}

For every data set, two distributions have been reconstructed differing in the level of smoothness and sparsity. The rational behind this is that the degree of sparsity as well as smoothness depends on the choice of $\beta$ as it will be shown later in this paper. This means that despite the usage of the proposed hyperparameter search more than a single pair of hyperparameters can be validly employed for MTGV regularization depending whether a smooth or sparse solution is preferred. In more detail, the proposed method chooses a $\beta$ close to the heel of the BRD curve, but the exact position of the heel is not clearly defined which gives the possibility for choosing a beta which is closer to the lower or the higher end of the heel allowing for different levels of smoothness and sparsity in the reconstructed distribution. For ease of discussion, to the solution with larger $\beta$ it will be referred as smooth and the solution with smaller $\beta$ will be described as sparse.
\begin{figure}[t!]
    \centering
    \begin{subfigure}[b]{0.32\textwidth}
        \centering
        \includegraphics[keepaspectratio, width=\textwidth]{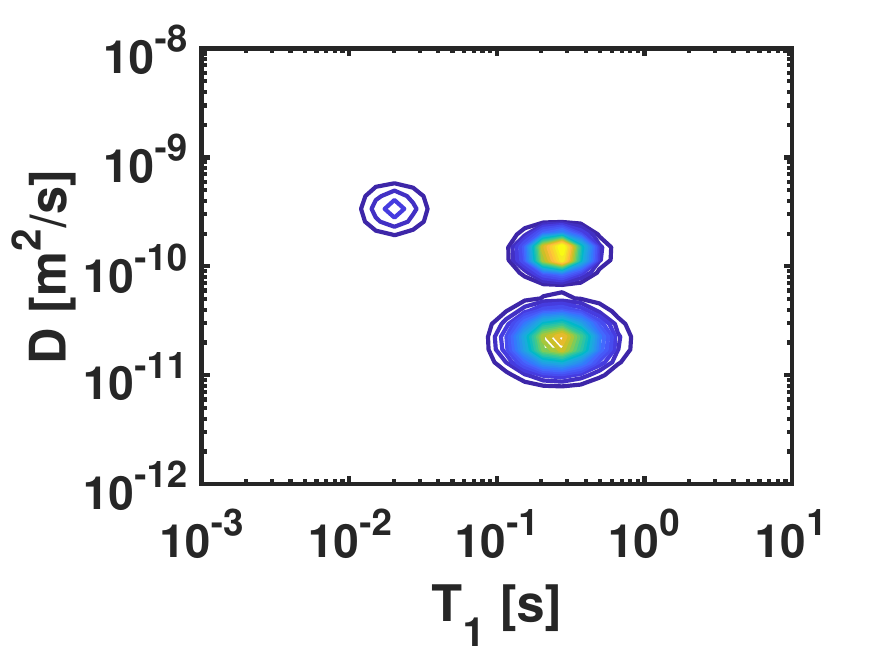}
        \caption{$T_1\mbox{-}D$~simulation}
        \vspace{1.1pc}
    \end{subfigure}
    \begin{subfigure}[b]{0.32\textwidth}
        \centering
        \includegraphics[keepaspectratio, width=\textwidth]{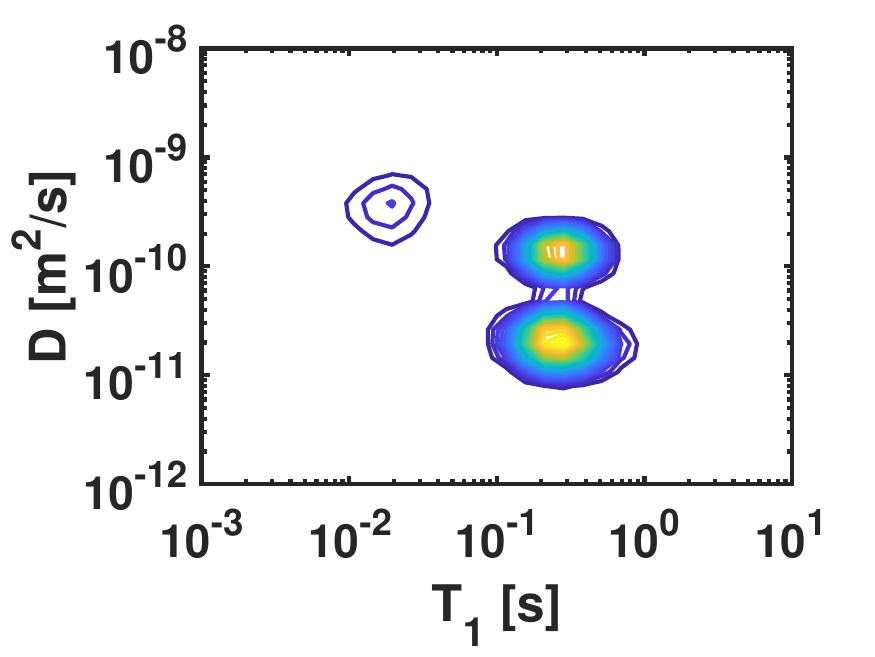}
        \caption{$T_1\mbox{-}D$~reconstruction ($\alpha = 8.4 \cdot 10^{3}$, $\beta = 3.9 \cdot 10^{-4}$)}
    \end{subfigure}
    \hspace{0.01\textwidth}
    \begin{subfigure}[b]{0.32\textwidth}
        \centering
        \includegraphics[keepaspectratio, width=\textwidth]{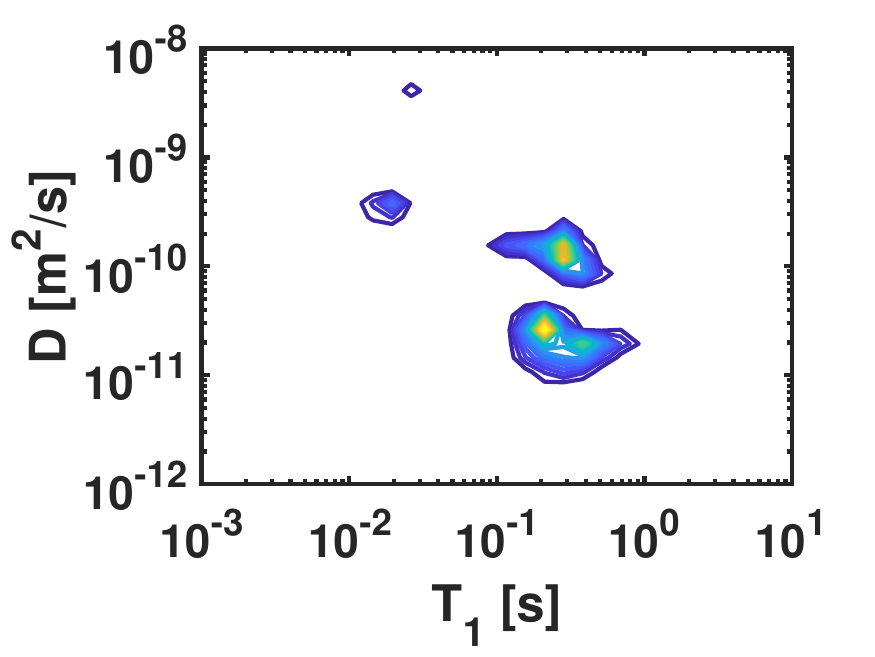}
        \caption{$T_1\mbox{-}D$~reconstruction ($\alpha = 3.4 \cdot 10^{3}$, $\beta = 3.4 \cdot 10^{-6}$)}
    \end{subfigure}
    \\
    \begin{subfigure}[b]{0.32\textwidth}
        \centering
        \includegraphics[keepaspectratio, width=\textwidth]{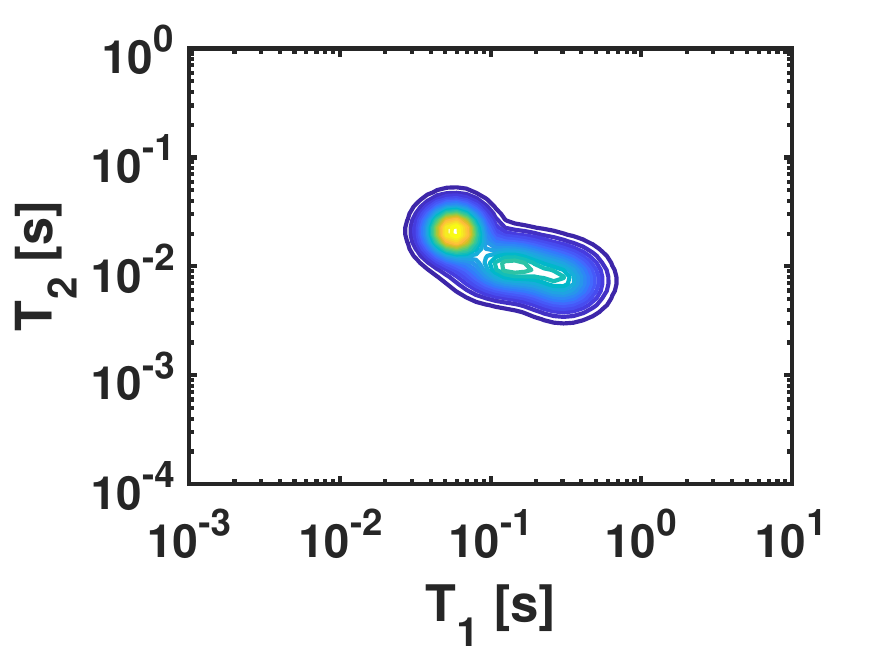}
        \caption{$T_1\mbox{-}T_2$~simulation}
        \vspace{1.1pc}
    \end{subfigure}
    \begin{subfigure}[b]{0.32\textwidth}
        \centering
        \includegraphics[keepaspectratio, width=\textwidth]{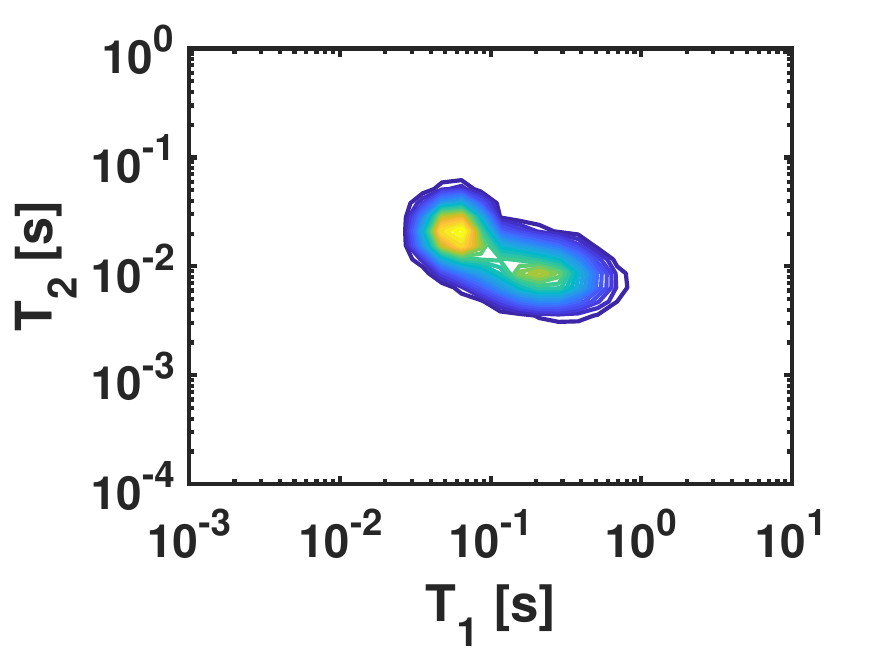}
        \caption{$T_1\mbox{-}T_2$~reconstruction ($\alpha = 3.2 \cdot 10^{4}$, $\beta = 6.3 \cdot 10^{-4}$)}
    \end{subfigure}
    \hspace{0.01\textwidth}
    \begin{subfigure}[b]{0.32\textwidth}
        \centering
        \includegraphics[keepaspectratio, width=\textwidth]{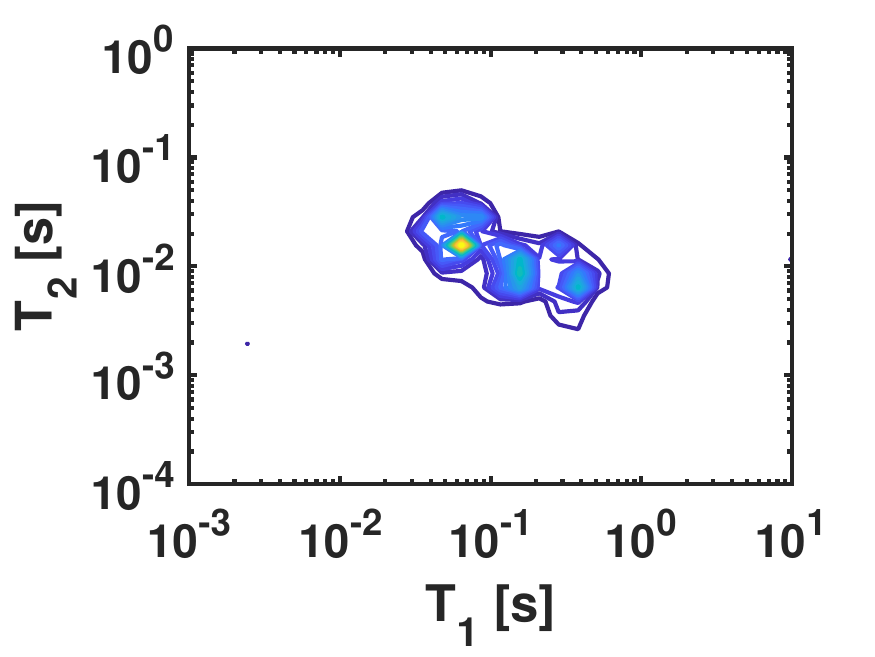}
        \caption{$T_1\mbox{-}T_2$~reconstruction ($\alpha = 2.9 \cdot 10^{4}$, $\beta = 2.8 \cdot 10^{-6}$)}
    \end{subfigure}
    \\
    \begin{subfigure}[b]{0.32\textwidth}
        \centering
        \includegraphics[keepaspectratio, width=\textwidth]{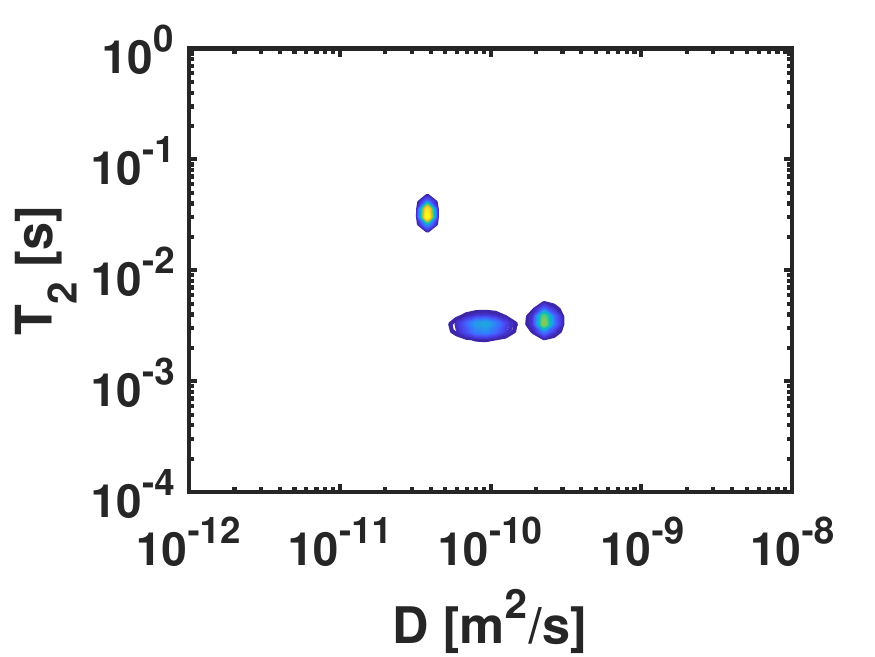}
        \caption{$D\mbox{-}T_2$~simulation}
        \vspace{1.1pc}
    \end{subfigure}
    \begin{subfigure}[b]{0.32\textwidth}
        \centering
        \includegraphics[keepaspectratio, width=\textwidth]{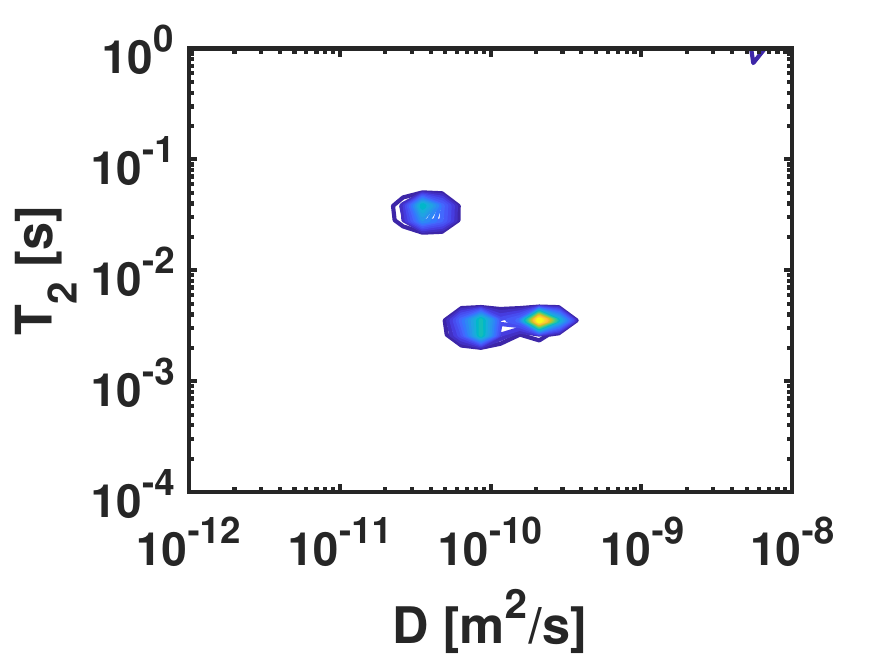}
        \caption{$D\mbox{-}T_2$~reconstruction ($\alpha = 1.6 \cdot 10^{4}$, $\beta = 5.5 \cdot 10^{-5}$)}
    \end{subfigure}
    \hspace{0.01\textwidth}
    \begin{subfigure}[b]{0.32\textwidth}
        \centering
        \includegraphics[keepaspectratio, width=\textwidth]{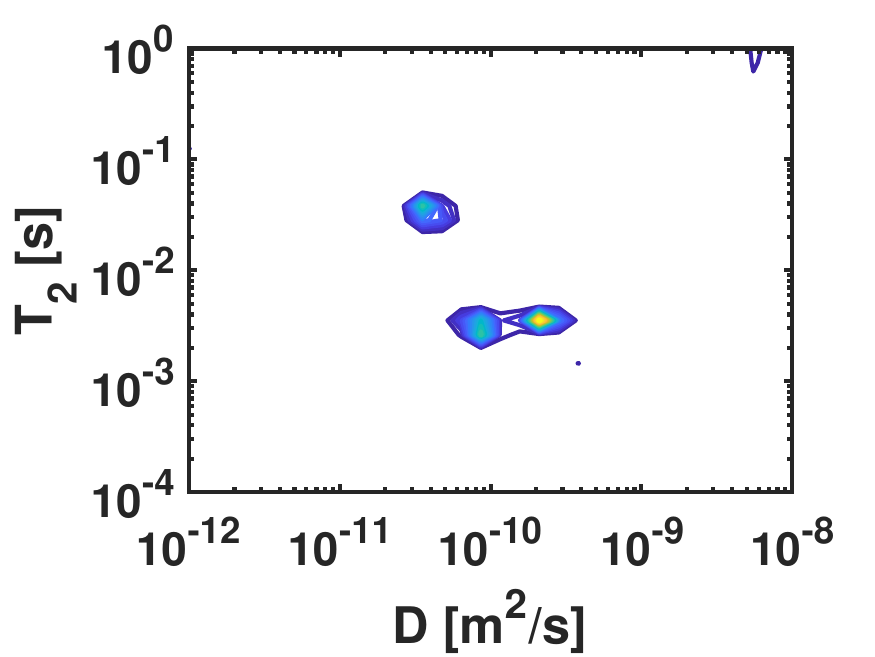}
        \caption{$D\mbox{-}T_2$~reconstruction ($\alpha = 1.7 \cdot 10^{4}$, $\beta = 7.5 \cdot 10^{-7}$)}
    \end{subfigure}
    \caption{Simulated (left), smoothly reconstructed (middle) and sparsely reconstructed (right) distributions.}
    \label{fig:sim_mtgv}
\end{figure}
In alignment with this, a full $\alpha$-search was conducted for the sparse as well as smooth solution, but for the purpose of computational efficiency only the values of $\alpha$, which fulfilled the convergence criterion were used to reconstruct a distribution. The original simulated distributions as well as both types of reconstructions are given in figure~\ref{fig:sim_mtgv}. The numerical values obtained for $\alpha$ and $\beta$ as a result of the proposed hyperparameter search are given in the figure captions of the reconstructed distributions. Comparing the simulated distributions with the results obtained from regularization, it becomes evident that for all investigated data sets the distribution was successfully reconstructed. 
\begin{figure}[t]
    \centering
    \begin{subfigure}[b]{0.31\textwidth}
        \centering
        \includegraphics[keepaspectratio, width=\textwidth]{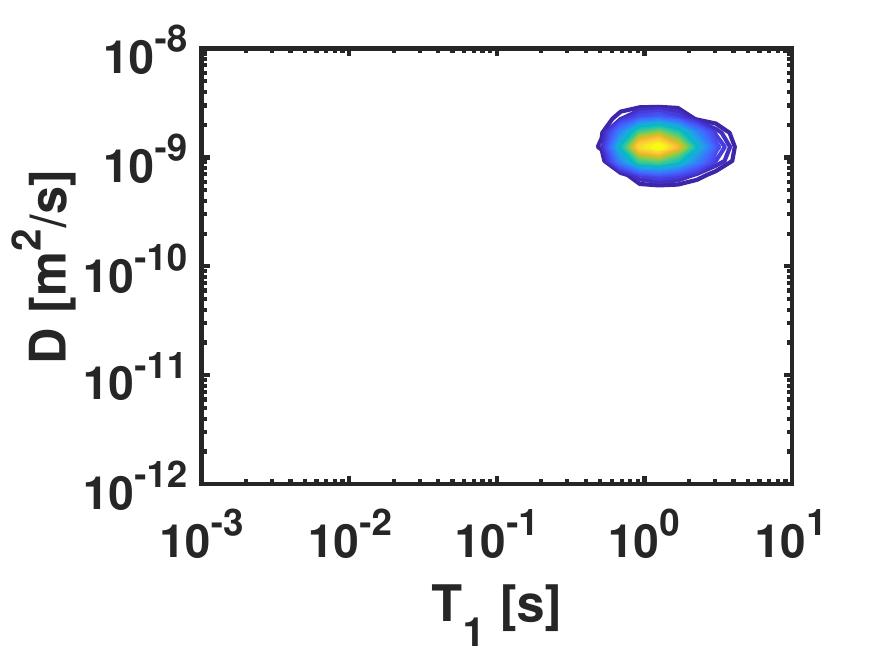}
        \caption{$T_1\mbox{-}D$~reconstruction ($\alpha = 7.5 \cdot 10^{1}$, $\beta = 5.3 \cdot 10^{-4}$)}
    \end{subfigure}
    \hspace{0.01\textwidth}
    \begin{subfigure}[b]{0.315\textwidth}
        \centering
        \includegraphics[keepaspectratio, width=\textwidth]{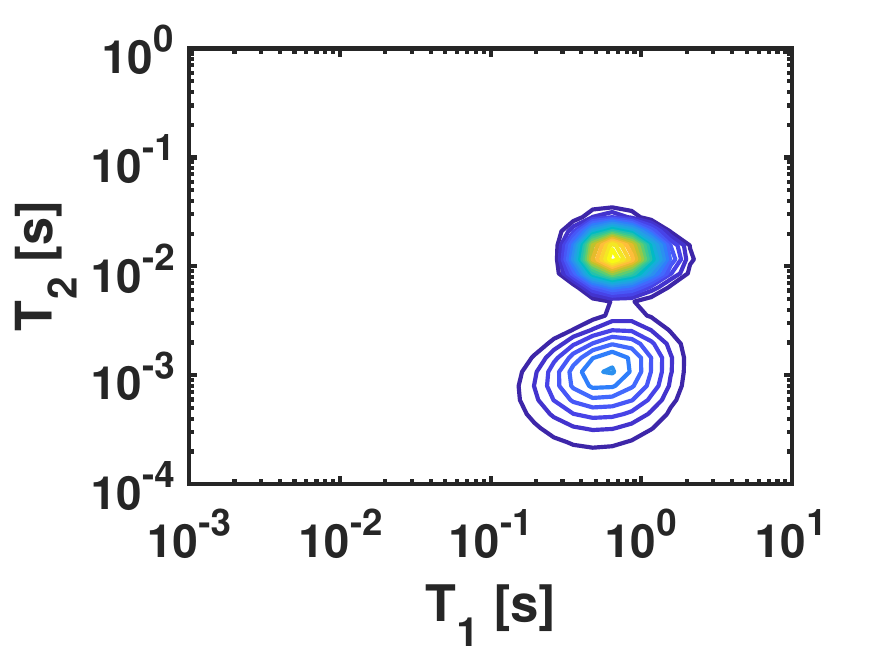}
        \caption{$T_1\mbox{-}T_2$~reconstruction ($\alpha = 2.1 \cdot 10^{0}$, $\beta = 3.7 \cdot 10^{-4}$)}
    \end{subfigure}
    \hspace{0.01\textwidth}
     \begin{subfigure}[b]{0.31\textwidth}
        \centering
        \includegraphics[keepaspectratio, width=\textwidth]{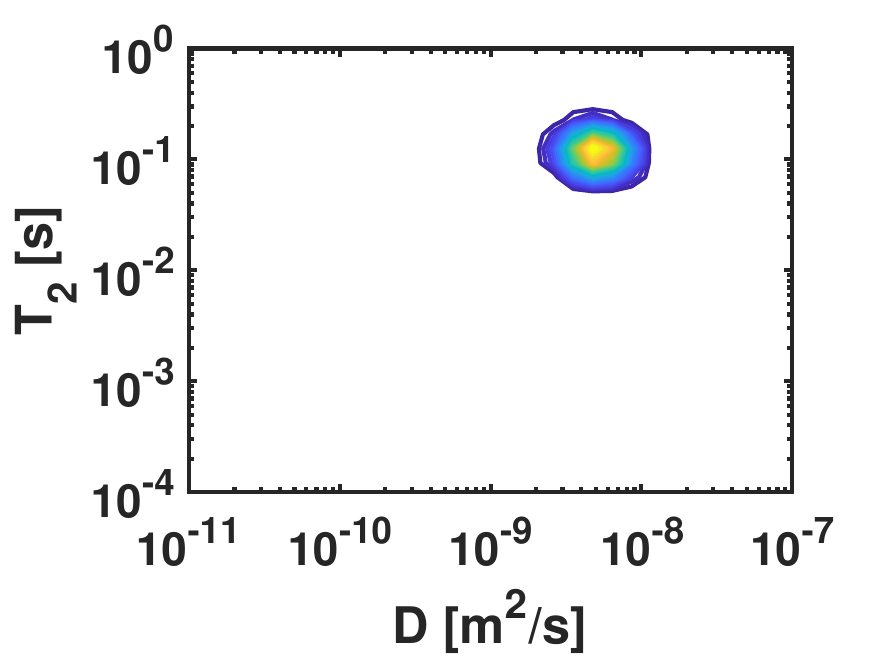}
        \caption{$D\mbox{-}T_2$~reconstruction ($\alpha = 7.1 \cdot 10^{0}$, $\beta = 4.5 \cdot 10^{-4}$)}
    \end{subfigure}
    \\
    \begin{subfigure}[b]{0.31\textwidth}
        \centering
        \includegraphics[keepaspectratio, width=\textwidth]{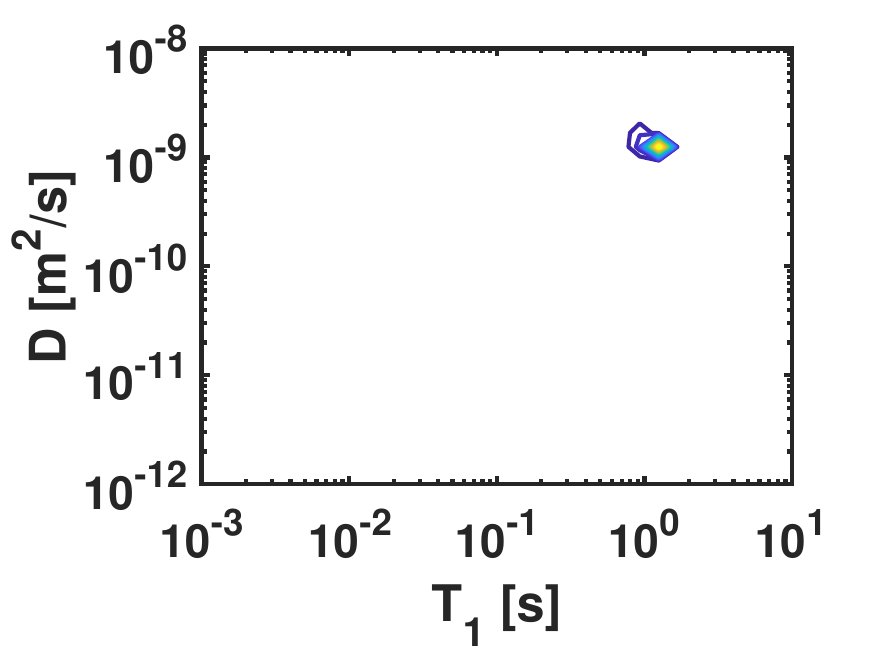}
        \caption{$T_1\mbox{-}D$~reconstruction ($\alpha = 4.3 \cdot 10^{2}$, $\beta = 3.5 \cdot 10^{-5}$)}
    \end{subfigure}
    \hspace{0.01\textwidth}
    \begin{subfigure}[b]{0.315\textwidth}
        \centering
        \includegraphics[keepaspectratio, width=\textwidth]{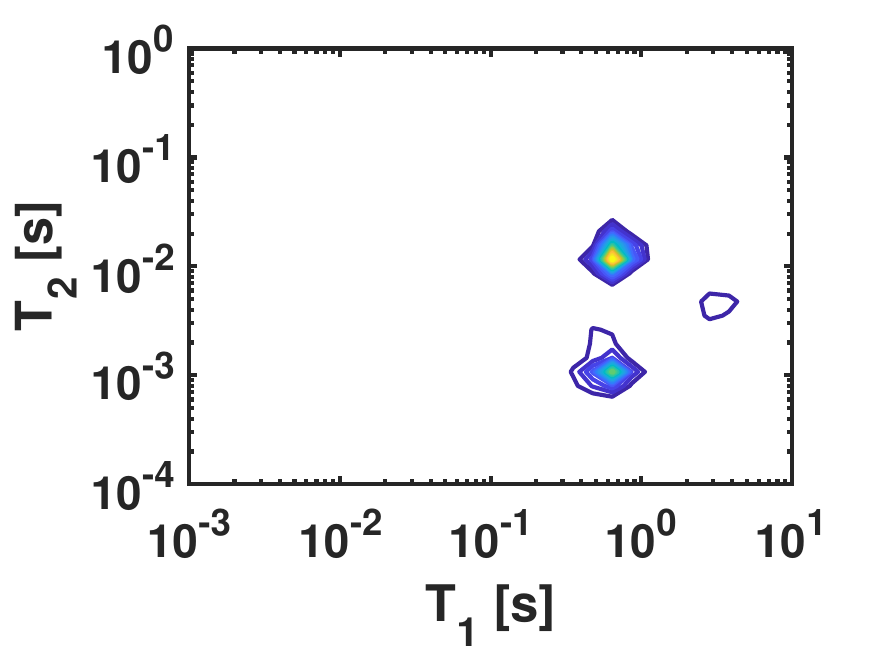}
        \caption{$T_1\mbox{-}T_2$~reconstruction ($\alpha = 5.5 \cdot 10^{0}$, $\beta = 1.9 \cdot 10^{-5}$)}
    \end{subfigure}
    \hspace{0.01\textwidth}
    \begin{subfigure}[b]{0.31\textwidth}
        \centering
        \includegraphics[keepaspectratio, width=\textwidth]{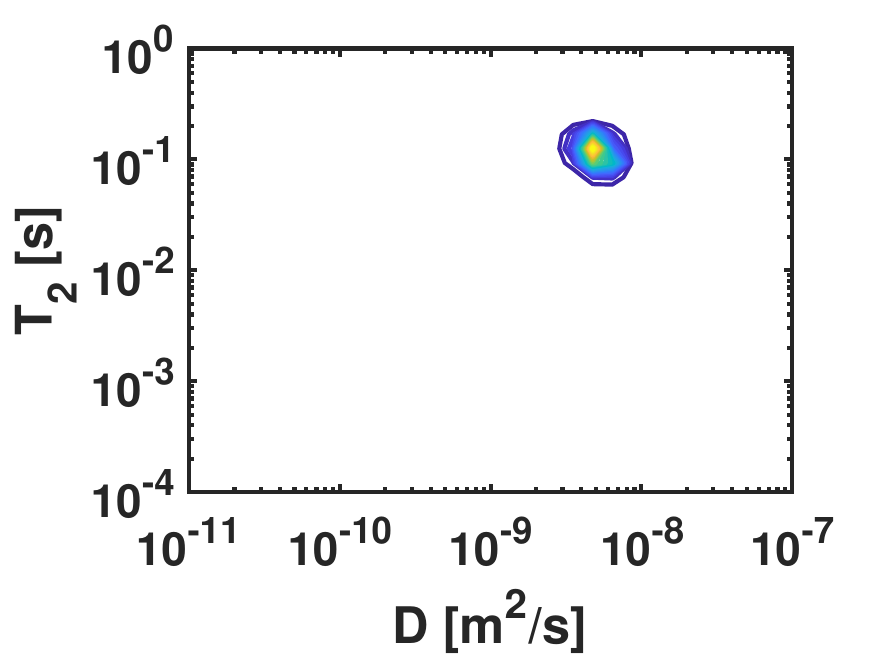}
        \caption{$D\mbox{-}T_2$~reconstruction ($\alpha = 5.5 \cdot 10^{0}$, $\beta = 3.1 \cdot 10^{-6}$)}
    \end{subfigure}
    \caption{Smooth (top) and sparse (bottom) reconstructions obtained from experimental NMR signals.}
    \label{fig:ex_mtgv}
\end{figure}
Focusing on the smooth solutions, figure~\ref{fig:sim_mtgv} shows that the choice of a larger $\beta$ provides a good reconstruction of the peak form especially if the underlying distribution is smooth and a small proneness to artifacts, but it suffers from poor resolution in the case of closely aligned peaks and from "over-smoothing" of sparse contributions. Comparing those results with the sparse reconstructions significant differences between both solutions become apparent. The sparse results show that selecting a smaller value for $\beta$ allows for a clearer separation of closely aligned components as well as a better reconstruction of sparse peaks with the cost that smooth sections are distorted and that regularization artefacts can be imposed. Overall, the smooth reconstructions can be considered closer to the original distribution but at the same time those solutions suffer from a poorer resolution of closely aligned or sparse peaks. Comparing the numerical values of $\beta$ for sparse \hl{as well as} smooth reconstructions, it becomes further evident, that with declining sparsity and consequently growing smoothness of the distributions, $\beta$ increases for both the sparse and the smooth solution. Those findings are in agreement with the theory outlined in section~\ref{sec:Theo} and~\ref{sec:Meth}. To recap briefly, $\beta$ controls the ratio between sparsity and smoothness of the reconstructed distributions, which means that smaller values of $\beta$ provide better solutions for sparse distributions and vice versa, which is exactly the observation made for the investigated data sets. 
\begin{figure}[t]
    \centering
    \begin{subfigure}[b]{0.48\textwidth}
        \centering
        \includegraphics[keepaspectratio, width=\textwidth]{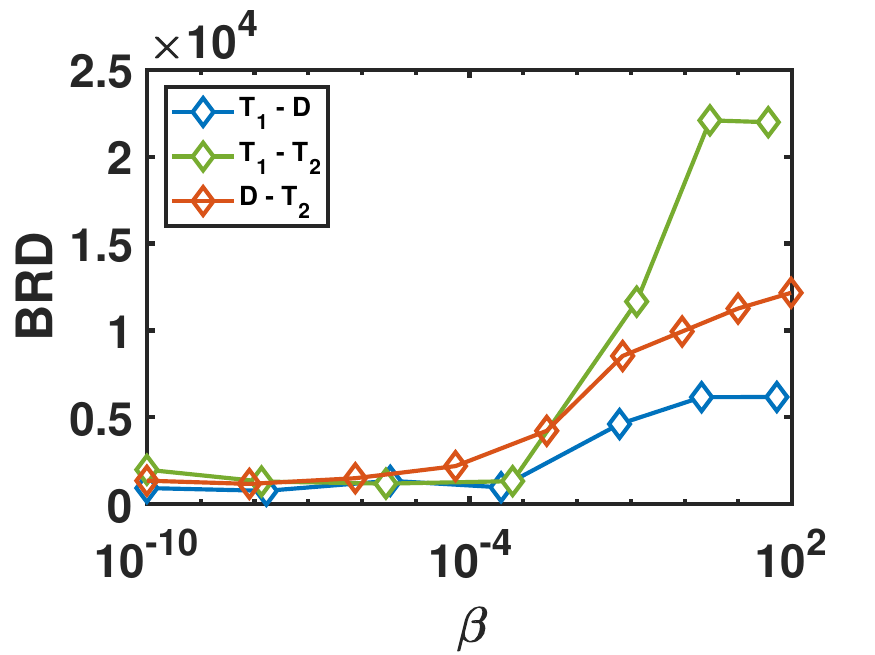}
        \caption{Simulation}
    \end{subfigure}
    \begin{subfigure}[b]{0.48\textwidth}
        \centering
        \includegraphics[keepaspectratio, width=\textwidth]{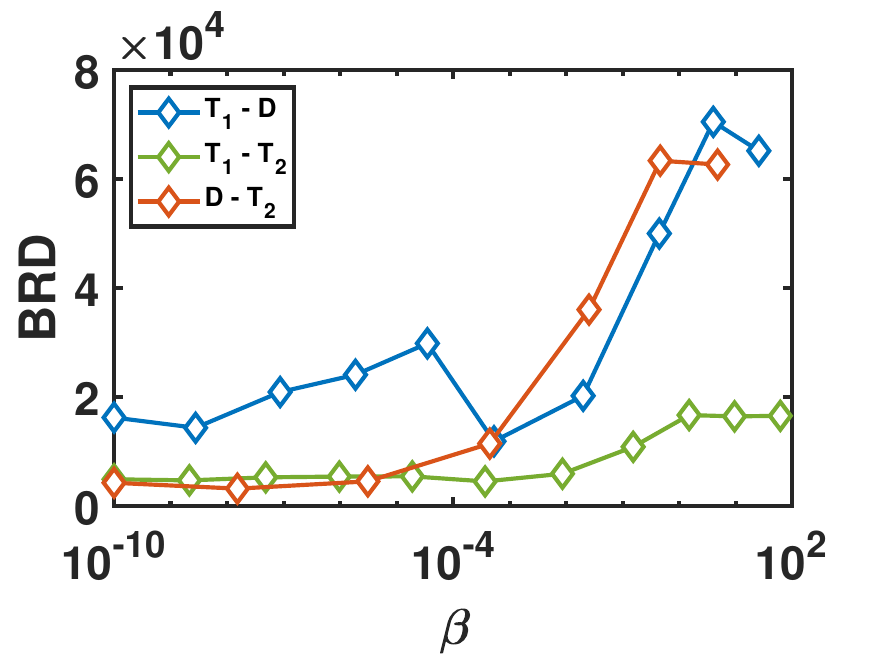}
        \caption{Experiment}
    \end{subfigure}
    \\
    \begin{subfigure}[b]{0.48\textwidth}
        \centering
        \includegraphics[keepaspectratio, width=\textwidth]{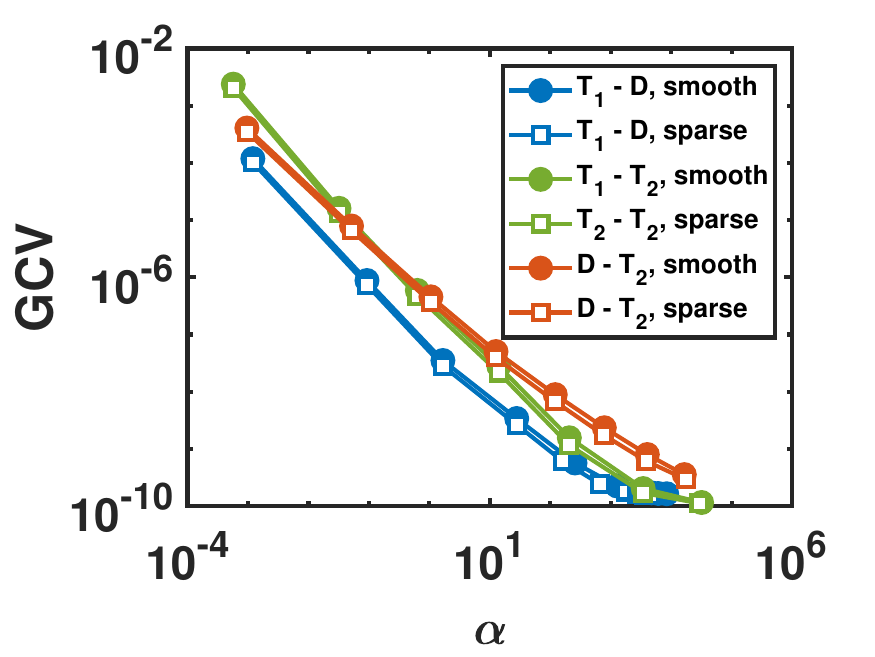}
        \caption{Simulation}
    \end{subfigure}
    \begin{subfigure}[b]{0.48\textwidth}
        \centering
        \includegraphics[keepaspectratio, width=\textwidth]{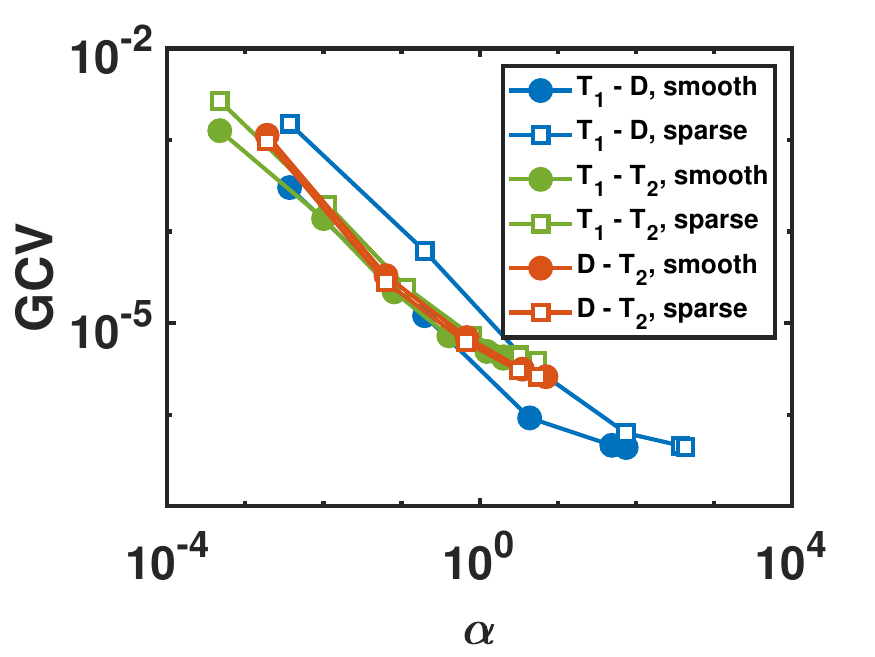}
        \caption{Experiment}
    \end{subfigure}
    \caption{BRD score (top) and GCV score (bottom) of the simulated (left) and experimental (right) data sets.}
    \label{fig:BRD_GCV}
\end{figure}
Focusing on the experimental data sets, it becomes evident that for all experiments a distribution could be successfully reconstructed independently of whether a smooth or sparse selection of $\beta$ was employed. The resulting reconstructions are given in figure~\ref{fig:ex_mtgv}. The differences between the sparse and smooth solution follow the same trends as it was discussed for the simulated data sets. Those results further show that in the case of experimental data, which means that usually the true distribution is unknown, an optimal hyperparameter selection cannot always be achieved on a pure numerical basis. For instance, the BRD scores of the smooth and sparse solution of the $T_1$-$T_2$-experiment differ only slightly, but represent very different distributions in terms of smoothness, sparsity and peak separation. This means, that in this or in similar cases, the optimal distribution can only be identified if further physical or chemical information, which allows differentiation between the distributions, is available. Consequently, the possibility to identify more than one set of optimal hyperparameters can be considered as a further advantage of the proposed method. \hl{To analyse the proposed hyperparameter search method in more detail,} the BRD score was plotted against $\beta$. The resulting figures for simulated as well as experimental data sets are depicted in figure~\ref{fig:BRD_GCV}. In the case of the $\alpha$-search, a plot of the GCV score against $\alpha$ is provided, but only the results obtained from the smooth and sparse solution of the $\beta$-search are shown. From figure~\ref{fig:BRD_GCV}, it becomes evident that in the numerical range studied, the BRD score converges to a lower bound or minimum with decreasing $\beta$ independent whether experimental or simulated data is used. In contrast, the inverse trend holds true for GCV. With increasing $\alpha$ the GCV score converges to a lower bound for experimental as well as simulated data sets. However, convergence alone does not ensure an efficient hyperparemter search, especially if a vast number of hyperparameters has to be explored before convergence is reached. In the case of the proposed method, only five to eleven different values for $\alpha$ or $\beta$ had to be tested until the convergence or stopping criterion was met. This can be considered as a very significant efficiency gain compared to the brute-force solution of looping through lists of hyperparameters, which in \hl{practice} can easily contain hundreds of different entries. In addition, the method architecture itself can be considered as another efficiency advantage. As outlined in section~\ref{sec:Meth}, the primal-dual algorithm is used twice during a full iteration cycle. Firstly, to obtain an optimal $\alpha$ and in a second instance to reconstruct the distribution. In agreement with Reci's initial work, the full reconstruction of a distribution requires between $10^3$ and $10^4$ iterations,\supercite{Reci2017RetainingExperiments.} whereas for all data sets tested in this paper, the estimation of $\alpha$ took ten or less iterations until the convergence criterion was met. From a viewpoint of computational efficiency, this means that the computational time used for the $\alpha$-search is negligible compared to the actual reconstruction of the distribution rendering the proposed method even more efficient. Overall, on a Windows $10$ desktop computer with $32$ GB of RAM and an AMD Ryzen $7$ $5800$X CPU the complete hyperparameter search including the full reconstruction of the distributions took less than three minutes independent of the data set used. In comparison, the brute-force method of looping through a list of hyperparameters can be easily one order of magnitude slower due to the necessity that for every $\alpha$-$\beta$-pair a full reconstruction has to be calculated. This further highlights the proposed method's efficiency gains as well as its high practicality for routine usage.
\section{Conclusion} 
\label{sec:Con}

In this paper, the concept of generalized cross-validation was applied to find an optimal $\alpha$ for the outlined MTGV regularization problem. It was further shown, that the MTGV cost function can be rearranged to a Tikhonov regularization problem for which a GCV score can be easily calculated and based on this an $\alpha$-update formula can be derived. This was combined with the Butler-Reeds-Dawson method, which was used to determine the second hyperparameter $\beta$. It was further shown, that in the investigated range of values both scores converge to a lower bound or a minimum, which allows to choose the hyperparameters accordingly. Overall, the proposed hyperparameter search allows to select $\alpha$ and $\beta$ efficiently, because only a small number of hyperparemeters have to be explored and therefore, it significantly increases the practicality for using MTGV regularization as a standard method for the inversion of NMR signals.

\printbibliography[title=References]

\end{document}